\documentclass{JHEP3}
\usepackage{graphicx,amsmath,amssymb}
\usepackage{epsfig,multicol}
\usepackage{epsf}
\usepackage{epstopdf}
\input{epsf.sty}
\usepackage{amssymb}
\usepackage{graphicx,amsmath,amssymb}

\newcommand{\ba}{\begin{eqnarray}}
\newcommand{\ea}{\end{eqnarray}}

\title{A Renormalization Group Approach to the Cosmological Constant Problem}
\author{S.-H. Henry Tye\footnote{Electronic mail: tye@lepp.cornell.edu} \\
\\
Laboratory for Elementary Particle Physics \\ Cornell University, 
Ithaca, NY 14853
}
\date{}

\abstract{In an earlier paper, it is proposed that, due to resonance tunneling effect, tunneling from a large cosmological constant $\Lambda$ site in the stringy comic landscape can be fast, while tunneling from a small $\Lambda$ site may take exponentially long time. Borrowing the renormalization group analysis of the conductance in the Anderson localization transition, we treat the landscape as a multi-dimensional random potential and find that the vastness of the landscape leads to a sharp transition at a small critical value $\Lambda_{c}$ from fast tunneling for $\Lambda > \Lambda_{c} $ to suppressed tunneling for $\Lambda_{c} > \Lambda >0$. Mobility in the landscape makes eternal inflation highly unlikely. As an illustration, we find that $\Lambda_{c}$ can easily be exponentially small compared to the string/Planck scale. These properties may help us in finding a qualitative understanding why today's dark energy is so small.}

\keywords{string theory, cosmic landscape, the cosmological constant problem, Anderson localization}

\begin{document}

\section{Introduction}

An outstanding problem in physics and cosmology is the smallness of today's dark energy value.
Treating it as a cosmological constant, it is many orders of magnitude smaller than the natural scale of gravity, namely, the Planck scale. Of course one can treat today's cosmological constant as a free parameter. However, dynamically it is difficult to understand why nature picks such a small value. So this is a naturalness problem (for some excellent reviews, see {\it e.g.},
Ref.\cite{Weinberg,Carroll:2000fy,Banks:2004zb,Copeland:2006wr,Polchinski:2006gy,Bousso:2007gp}).
In string theory, the natural energy scale is the string scale. Again, today's cosmological constant is many orders of magnitude smaller than the string scale. 
So, at first sight, the naturalness problem persists. 
Recent understanding of string theory in flux compactification \cite{Giddings:2001yu,Kachru:2003aw} suggests that there are many classically stable vacuum solutions. 
The stringy cosmic landscape is vast and complicated \cite{Bousso:2000xa,Kachru:2003aw,Susskind:2003kw,Douglas:2003um,Denef:2004cf}. It is argued in Ref.\cite{Bousso:2000xa} that it is the vastness of the landscape that allows a classically stable vacuum with such a small cosmological constant. It is argued in Ref.\cite{Tye:2006tg} that it is precisely the vastness of this landscape that may provide a partial solution to the naturalness problem. Here, the latter argument is supplemented with a more analytical investigation.
 
There are exponentially (if not infinitely) many classically stable vacua in string theory. Due to tunneling effects, ones with positive vacuum energies are meta-stable. This landscape is described by a set of 
scalar modes, namely, light closed string modes known as moduli and positions of branes and fluxes.
(We shall refer to them collectively as moduli.)
Depending on the location of any specific site, the number $d$ of moduli in the stringy landscape describing its neighborhood may vary from dozens to hundreds. So a large $d$ parameterizes the vastness of the landscape. In Ref.\cite{Tye:2006tg}, it is argued that resonance tunneling effect can enable a fast decay of a site with a string scale vacuum energy $\Lambda$ in the stringy cosmic landscape.
The key point may be illustrated with the following simplistic example.

Let the tunneling rate from one site to any of its neighboring site be $\Gamma_{0}$, which is dictated by the minimum (not typical) barrier tunneling path. 
Tunneling from a positive $\Lambda$ site to a larger $\Lambda$ site (with smaller entropy) takes a very long time and is ignored here.
Note that tunneling from a positive $\Lambda$ site to a negative $\Lambda$ site is also ignored for at least one of the following reasons : (1) it takes an exceedingly long time, (2) it leads to a big crunch \cite{Coleman:1980aw}, (3) it pops right back up \cite{Banks:2005ru,Aguirre:2006ap}. 
In this sense, $\Lambda=0$ is special.

Consider tunneling in a $d$-dimensional hyper-cubic lattice.
Ignoring resonance tunneling for the moment, then the total tunneling rate is given by
\ba
\label{nrt}
\Gamma^{nr}_{t} \sim 2 d\ \Gamma_{0}
\ea
since there are $2d$ nearest neighbors.
Here, $\Gamma^{nr}_{t}$ grows linearly with $d$.
Since $\Gamma_{0}$ is exponentially small, $\Gamma^{nr}_{t}$ is also exponentially small.
Now let us include the resonance/efficient tunneling effect (see the appendices for a brief review and further discussions). For a generic wavefunction at a site,
its tunneling rate to its next-to-neighboring sites will have comparable $\Gamma_{0}$, 
not $\Gamma_{0}^{2}$, as naively expected. 
Since the number of paths for resonance tunneling increase like the volume, it is argued that the effective total tunneling rate goes like 
\ba
\label{rst}
\Gamma_{t} \sim  n^{d}\ \Gamma_{0}
\ea
where $n$ signifies the effective number of steps where resonance/efficient tunneling acts with tunneling rate not much suppressed with respect to $\Gamma_{0}$. That is, as a function of $d$, 
$\Gamma_{t}$ actually grows exponentially, not linearly. Now it becomes possible that 
$\Gamma_{t}$ can be of order unity for large enough $d$ and $n$. 

The scenario then goes as follows. Tunneling will be fast when the site has a relatively large $\Lambda$, since it has many paths and nearby sites to tunnel to. This fast tunneling process can happen repeatedly, until it reaches a site with $\Lambda$ smaller than some critical value 
$\Lambda_{c}$. Most if not all nearby sites of this low $\Lambda$ site have larger $\Lambda$s, 
so it will not tunnel to. That is, the effective $n \rightarrow 1$ in most directions
so its total decay width becomes exponentially small. This low $\Lambda$ site may describe today's universe.
For this scenario to work, such a shut off of fast tunneling must be sharp, 
so that the universe never enters eternal inflation. 
Otherwise, the universe would arrive at a meta-stable site with some intermediate $\Lambda$, and have enough time to expand away the radiation/matter present and enter into eternal inflation 
{\footnote{ In brane inflation, eternal inflation of the random walk type generically does not take place \cite{Chen:2006hs}. So here we are concerned only with eternal inflation of the tunneling type in the landscape.}}. 
If this happens, some (most) parts of our universe would still be in an eternally inflationary phase today. To avoid this situation, the transition from fast to (exponentially) slow tunneling must be sharp.  

 In a realistic situation, the stringy cosmic landscape is 
 very complicated. In this paper, I shall treat the landscape in the Schr${\ddot o}$dinger approach, with the landscape looking like a random potential. Using the scaling theory developed in condensed matter physics, I argue that the transition from fast tunneling to exponentially slow tunneling is sharp, and happens at an exponentially small $\Lambda_{c}$. 

Let $\Psi (a, ..., \varphi_{j},  \phi_{i}, ... )$ be the wavefunction of the universe in the landscape, where $a$ is the cosmic scale factor, $\varphi_{j}$ are the values of the closed string moduli and $\phi_{i}$ are the positions of the $D$-branes (and fluxes) in the compactified bulk (which can decompactify). In general, the $\phi_{i}$ and the  $\varphi_{j}$ are all coupled, so movements of the branes will shift the values of the moduli. Suppose, in some situation, this wavefunction can be approximately described by the positions of the $D$3-branes only. Then fast tunneling becomes even more likely since it has been shown recently that DBI tunneling, though still exponentially suppressed, can be exponentially more efficient than tunneling as described by the Coleman-DeLucia mechanism \cite{Brown:2007ce}. 

As we shall see, there is a sharp transition from fast tunneling to exponentially long tunneling, when the wavefunction is trapped or localized. 
At the transition, there is a correlation length that blows up, and one can use the renormalization group to study its scaling behavior. A number of properties that are relevant to the questions we like to address are those of a specific universal class and so are independent of the details of the landscape. In this sense, we can say something concrete. This is crucial here, since our knowledge of the landscape is extremely limited. Since the landscape is very complicated and, at first sight, it may look like a random potential in $d$ dimensions, the problem is analogous to a random potential problem in condensed matter physics. Fast tunneling means mobility, or conducting (metalic), while strongly exponentially suppressed tunneling means localization, or insulating. 
The later is known as the Anderson localization \cite{Anderson}. The transition between the conducting and the insulating phases has scaling properties, allowing the application of renormalization group techniques \cite{Wilson} to understand some of the universal properties. 

Here I like to borrow the scaling theory of Anderson localization transition developed by Abrahams, Anderson, Licciardello and Ramakrishnan \cite{AALR} and others \cite{Shapiro,Lee,Sadovskii} in a random potential to re-examine the validity of the above 2 key features necessary for the solution to the cosmological constant problem in the cosmic landscape. 
The renormalization group treatment in this problem is phenomenological. Fortunately, they have been checked to a large extent by experiments and additional analyses for $d=3$ (see e.g., \cite{Lee,Sadovskii}). 
The conductivity is finite for a conductor, and is zero for an insulator. 
Here, we are interested in the behavior of the conductivity, or more appropriately as it turns out, the conductance $g$ of the system. Expressed in terms of the natural unit of the system, $g$ is dimensionless. The scaling theory shows that there is a transition for $d>2$, and the transition takes place at a critical value of $g=g_{c}$, where the corresponding renormalization group flow $\beta$-function vanishes, i.e.,  $\beta (g_{c})=0$.
We shall see that $g_{c}(d)$ decreases exponentially as a function of $d$. That is, for large $d$, an exponentially small conductance can lead to mobility. 
Also, $g_{c}$ is an unstable fixed point, implying that the transition between mobility and 
localization is a sharp one.

Consider a specific (enveloping) wavefunction $\psi ({\bf r}) \sim e^{-|{\bf r}|/\xi}$ at site ${\bf r}={\bf 0}$ with a typical $\Lambda$ in the cosmic landscape (see Figure 1(b)). The wavefunction seems to be completely localized at the site if the localization length $\xi \ll  a$, where $\xi$ measures the size of the site and $a$ is the typical spacing between sites. They are determined by local properties and $\xi$ is known as the localization length. 
With resonance tunneling, $\xi$ can be somewhat bigger, although $\psi ({\bf r})$ may still be exponentially suppressed at distance $a$. 
At distance scales around $a$, the conductance goes like
\ba
\label{ga}
g(a) \sim |\psi(a)| \sim e^{-a/\xi}
\ea
Note that, for $a \gg \xi$, $\Gamma_{0} \sim |\psi(a)|^{2} \sim e^{-2a/\xi}$. 

Given $g=g(a)$ (\ref{ga}) at scale $a$, what happens to $g$ at distance scales $L \gg a$ ? That is, how does $g$ scale ? The scaling theory of $g$ is well studied in condensed matter physics.
It turns out that there is a critical conductance $g_{c}$ which is a function of $d$.
If $g(a) < g_{c}$, then $g(L) \sim e^{-L/\xi} \rightarrow 0$ as $L \rightarrow \infty$ and the landscape is an insulating medium. The wavefunction is truly localized. 
Tunneling will take exponentially long and eternal inflation comes into play. 
If $g(a) > g_{c}$, the conductivity is finite  ($g(L) \sim (L/a)^{(d-2)}$) as $L \rightarrow \infty$ and the landscape is in a conducting/mobile phase. The wavefunction is free to move. Around the transition point, there is a correlation length that blows up at the phase transition. We shall see that, at length scale $a$, this correlation length can be identified with $\xi$. That is, $\xi$ plays the role of the correlation length. This is how, given $g(a)$ (\ref{ga}) at scale $a$, $g$ can grow large for large $L$. 

In this paper, we argue that, for large $d$, the critical conductance is exponentially 
small, $$g_{c} \simeq e^{- (d-1)}$$
So it is not difficult for $g(a) > g_{c}$; given $g(a)$ (\ref{ga}), we see that fast tunneling happens when $\Gamma_{0} > e^{-2(d-1)}$, or 
\ba
d > \frac{a}{\xi} +1
\label{condition}
\ea
so the wavefunction of the universe can move freely in the quantum landscape even for an exponentially small $g(a)$.
Since the $\beta$-function has a positive slope at the localization transition point (implying an unstable fixed point), the transition between the insulating (localization) phase and the conducting (mobile) phase is sharp. 
We see that it is precisely the vastness of the cosmic landscape (as parameterized by a large $d$) that allows mobility even if the wavefunction is localized. This mobility allows the wavefunction to sample the landscape very quickly. It also means a semi-classical description of the landscape is inadequate.
 
In the mobility phase, the wavefunction and/or the $D$3-branes (and moduli) move down the landscape to a site with a smaller $\Lambda$. At that site, some of its neighboring sites have larger $\Lambda$s, so the effective $a$ increases, leading to a larger effective critical conductance $g_{c}$. This happens until the condition (\ref{condition}) is no longer satisfied and the $D$3-brane is stuck at a small $\Lambda$ site. This transition takes place at a critical $\Lambda_{c}$. 
The mobility shuts off for $\Lambda < \Lambda_{c}$, and the universe ends up at a site with a small 
$\Lambda < \Lambda_{c}$.  This suggests that the quantum landscape is a mixture of two components : the sites composed of $\Lambda >\Lambda_{c}$ form a conducting medium while the sites composed of 
$\Lambda <\Lambda_{c}$ form an insulating medium. 
In terms of the string scale, we intuitively expect $\Lambda_{c}$ to be exponentially small. To explain the cosmological constant problem, we require $\Lambda_{c}$ to be larger than, but not too many orders of magnitude larger than today's dark energy.
A more detailed analysis of this quantum landscape is necessary to find the critical value 
$\Lambda_{c}$. As an illustration, we give a simple toy-like scenario, where we find that $\Lambda_{c}$ can easily be exponentially small compared to the string or the Planck scale. For a flat distribution of number of sites with the 4-dimensional cosmological constant between the string scale $M_{s}$ and zero, we see that the critical cosmological constant goes like $d^{-d}M_{s}^{4}$. This yields a critical cosmological constant at the right order of magnitude for $d \sim 50$, a reasonable value. Clearly a better estimate will be valuable.

The time for one e-fold of inflation is Hubble time $1/H$ ($H$ is the Hubble parameter) while the lifetime of the site is naively given by $(H \Gamma^{nr}_{t})^{-1}$. Since $\Gamma^{nr}_{t}$ (\ref{nrt}) is exponentially small, eternal inflation seems unavoidable. 
 However, in the quantum landscape, 
we find that the mobility time scale is much shorter than the Hubble time scale; so we conclude that eternal inflation in the landscape is highly unlikely. The cosmological evolution of the universe remains to be understood. The quantum landscape does open new doors that should be explored. Some speculations are discussed. Analogous to condensed matter physics where the transition can go from localization to superconductivity, one may wonder if the landscape is not only conducting but superconducting, and what are its implications. 

An upcoming paper \cite{Sarangi:2007jb} comes to similar conclusions on fast tunneling as in Ref.\cite{Tye:2006tg} in a different approach. See Ref.\cite{Kane:2004ct,Kobakhidze:2004gm,MersiniHoughton:2005im,She:2007by,Podolsky:2007vg} for some other related discussions. 

In Section 2, we review the general scaling theory of Anderson transition and apply it to the landscape. In Section 3, we consider a specific model due to Shapiro \cite{Shapiro}. We find that the critical conductance is exponentially small for large $d$ and the condition is given in Eq.(\ref{condition}). In Section 4, we try to extract the critical $\Lambda_{c}$ from the critical conductance. As an illustration, we show that $\Lambda_{c}$ can easily be exponentially small compared to the string or the Planck scale. 
Section 5 contains some discussions concerning the overall scenario. We explain how eternal inflation is avoided and give some speculative overview of the scenarios we have in mind. Section 6 contains some final remarks. Appendix A reviews very briefly resonance tunneling, efficient tunneling and fast tunneling used in Ref.\cite{Tye:2006tg}. Appendix B discusses the meaning of the resonance tunneling 
effect in intuitive terms, which should apply in both quantum mechanics and quantum field theory. 
Appendix C interprets the Hawking-Moss tunneling formula in view of the resonance tunneling effect.

\section{The Scaling Theory of the Conductance Transition}

Consider a particle moving in a discordered (random potential) medium. 
The phase of its wavefunction varies randomly. The distance over which it fluctuates by $2\pi$ defines the mean free path $l$, the microscopic length scale of interest here.
When the disorder is small, the particle is scattered randomly and its wavefunction usually
changes at the scale of the order $l$.
However, the wavefunction remains extended plane-wave-like (Bloch wave-like) through the
medium. If the wavefunction is initially localized at some site, it would quickly spread and move towards low potential sites. This mobility implies that the system is a ``conductor''.

\begin{figure}[ht]
\begin{center}
\leavevmode
\includegraphics[width=0.8\textwidth,angle=0]{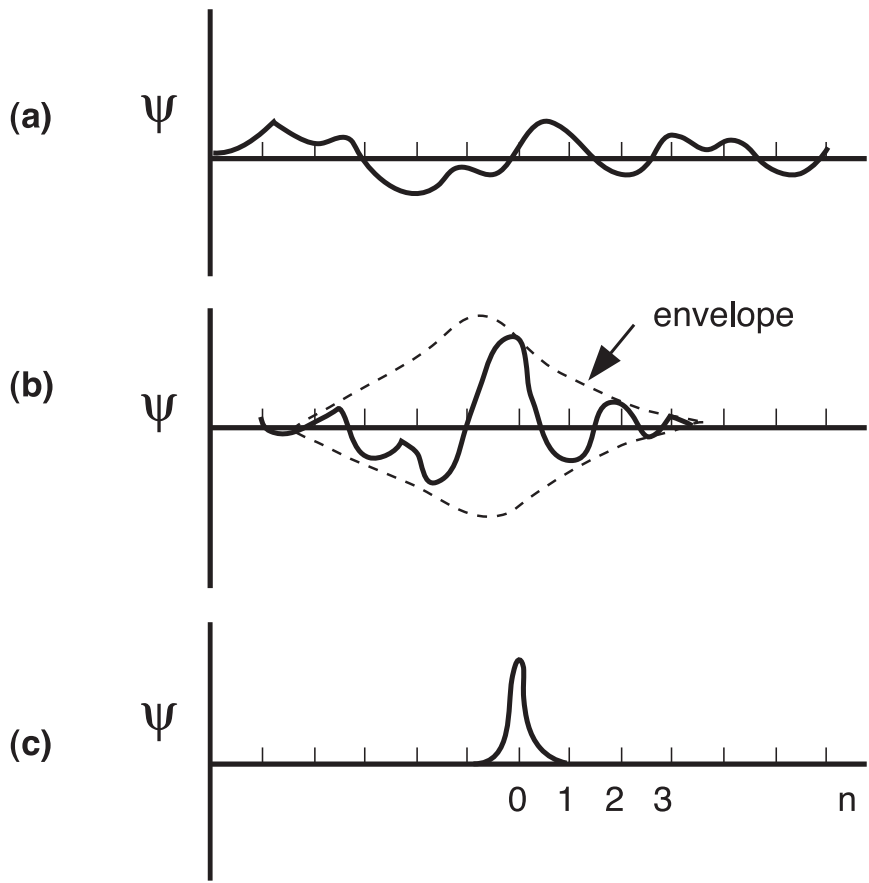} 
\caption[]{Schematic sketches of the wavefunction $\psi$ in the landscape : (a) an extended state, with mean free path $l \gg a$, where $a$ is the spacing between sites, (b) a weakly localized state, with size $\xi > a$, where the envelope is described by $|\psi({\bf r})|$ in Eq.(\ref{wavef1}), and (c) a localized state, with $\xi <  a$. Here $n$ is the number of lattice spacing of size $a$ away from ${\bf r} = {\bf 0}$. 
\label{fig1}
}
\end{center}
\end{figure}

Anderson \cite{Anderson} showed that the wave function of a quantum
particle in a random potential can qualitatively change its nature if the
randomness becomes large enough (strongly disordered). 
In this case, the wavefunction becomes localized, so that its
amplitude (envelope) drops exponentially with distance from the center of localization ${\bf r}_{0}$:
\begin{equation}
\label{wavef1}
|\psi({\bf r})|\sim \exp(- |{\bf r-r}_{0}|/\xi)
\end{equation}
where $\xi$ is the localization length. This situation is shown qualitatively
in Figure \ref{fig1}.
When the particle wavefunction is completely localized at a single site, $\xi$ approaches a value comparable to the typical spacing $a$ between sites. In $d \le 3$, it will take an exponentially long time to tunnel to a neighboring site. The lack of mobility implies that the system is an ``insulator''. 

The physical meaning of Anderson localization is relatively simple:  coherent
tunneling of particles is possible only between energy levels with the same energy. 
However, in case of strong randomness, $l \rightarrow a$ and the states with the same
energy are too far apart in space for tunneling to be effective. Since we are interested in large $d$, while condensed matter physics systems typically have $d \le 3$, we shall introduce a slightly refined definition here : \\
(1) an extended state if $\xi \gg a$, as shown in Figure \ref{fig1}(a);   \\
(2) a weakly localized state if $\xi \gtrsim a$, as shown in Figure \ref{fig1}(b); \\
(3) a localized state if $\xi < a$, as shown in Figure \ref{fig1}(c); \\
(4) a strongly localized state when $\xi \ll a$. As we shall see, for large $d$, a strongly localized state can still lead to mobility if it satisfies the condition (\ref{condition}). When this condition is not satisfied, we have \\
(5) a truly localized state. \\
For $d \le 3$, a weakly localized state will have lost mobility already (although it is believed that, under the right conditions, superconductivity can take place in the localization region). That a strongly localized state may still lead to mobility for large $d$ is not surprising. For large $d$, the particle has many directions for coherent or resonant tunneling. In a random potential environment, it is much more likely that some direction allows easy motion. The goal is to quantify this intuition. We shall continue to use the language of condensed matter physics : $a$ is the microscopic scale while $L$ is the macroscopic scale. 
 
For fixed $d$, we expect a transition from a ``conductor'' to an ``insulator'' as the potential barriers rise
and fast tunneling is suppressed. To learn more about this transition, we should focus on the behavior of the conductivity, or, more appropriately, the conductance. 
It turns out that the behavior of such a transition can be described by a scaling theory similar to that used in the theory of critical phenomena \cite{Wilson}. 
In the scaling theory of the transition between mobility and localization, one considers the behavior of the conductance as a function of the sample size $L$. Dimensionally, we shall choose some microscopic units (such as the typical spacing $a$ between neighboring sites or the string scale 
$\alpha '$) so that the conductance (the Thouless number) $g(L)$ is dimensionless.

The main physical idea of this approach is based upon a series of scale transformations
from smaller to larger ``cells'' in coordinate space
with appropriate description of the system by transformed
parameters of the Hamiltonian. These transformations
form the so called renormalization group. Here the localization length plays the role of the
correlation length as we approach the transition point from the 
insulator side. 
It implies that the transition is relatively insensitive to local details, which is crucial 
since we do not know the details in the neighborhood of a typical site in the cosmic landscape. 

Following  Ref.\cite{AALR}, localization is described in terms
of  the conductance $g$ as a function of a sample size $L$. 
If the states are localized, conductivity is zero (or exponentially small) and matrix elements for transitions between different states drop exponentially on distances of the order of $\xi$.  Then one expects that for $L\gg \xi$,\ the effective conductance becomes exponentially small:
\begin{equation}
g_{d}(L) \sim e^{-L/\xi}
\label{gloc}
\end{equation}
For a small disorder, the medium is in a metallic state and the
conductivity $\sigma$ is independent of the sample size if the size is much larger than the mean
free path, $L\gg l$.
Conductance is determined in this case just by the usual Ohm's law and for a
$d$-dimensional hypercube, we have:
\begin{equation}
 g_{d}(L)=\sigma L^{d-2}
\label{gconduc}
\end{equation}
based on a simple dimensional argument.
(Recall that in $d=3$, $g = \sigma $(Area)$/L \sim \sigma L$.) 
The conductance $g=g(a)$ at length scale $a$ is a microscopic measure of the disorder. We see that $g(L)$ may end up with one of the 2 very different asymptotic forms for $L\gg a$. Here $\sigma$ is rescaled so that $g$ is dimensionless.

\begin{figure}[ht]
\begin{center}
\leavevmode
 \includegraphics[width=0.9\textwidth,angle=0]{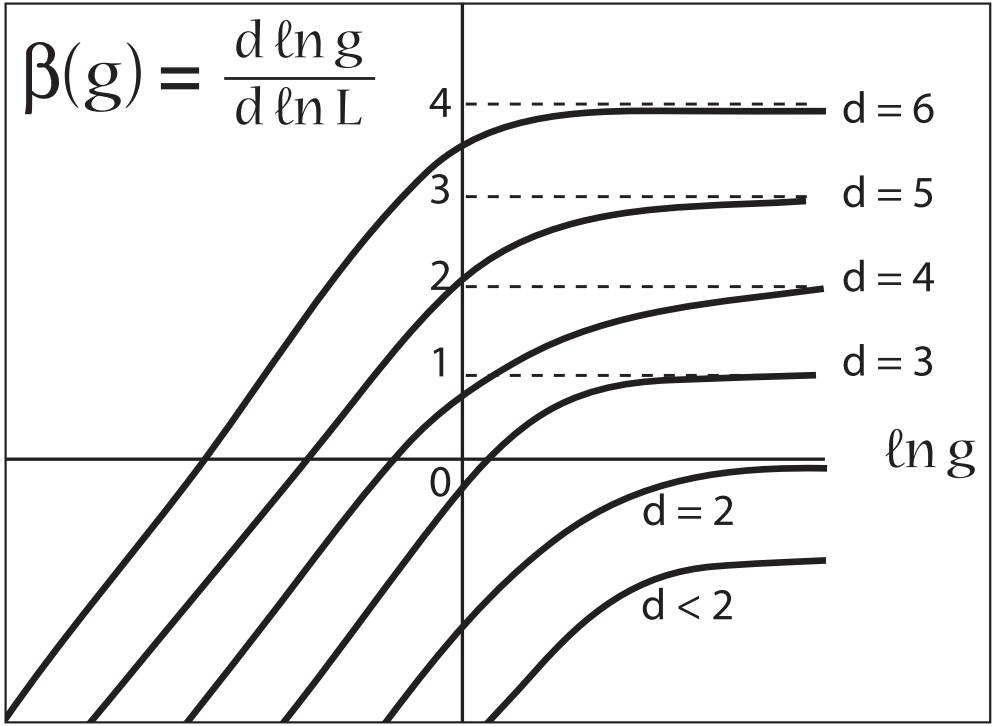} 
\caption{The $\beta$-function of the dimensionless conductance $g$, $\beta_{d}(g)$ versus $\ln g$, for different values of $g$ and $d$. 
The cases for $d<2$, $d=2$ and $d=3, 4, 5, 6$ are shown. For $g \rightarrow 0$, $\beta_{d}(g)$ is linear in $\ln g$, and $\beta_{d}(g) \rightarrow d-2$ as $g \rightarrow \infty$. 
There is no fixed point for $d<2$. For $d>2$, there is always a fixed point (the zero of the $\beta$-function at $g=g_{c}(d)$) and the slope at the fixed point is positive. For large $d$, the spacings of the fixed points for adjacent $d$ values are approximately equal in $\ln g$. 
\label{fig2}}
\end{center}
\end{figure}

Elementary scaling theory of localization assumes that $g$ of a 
$d$-dimensional hypercube of size $L$ satisfies the simplest differential equation of a
renormalization group, where 
\begin{equation}
\beta_{d}(g_{d}(L))=\frac{d \ln g_{d}(L)}{d \ln L}
\label{betagL}
\end{equation}
that is, the $\beta$-function $\beta_{d}(g)$ depends only on the
dimensionless conductance $g_{d}(L)$. Then the qualitative behavior of $\beta_{d}(g)$ can be
analyzed in a simple way by interpolating it between the 2 limiting forms given by Eq.(\ref{gloc}) and Eq.(\ref{gconduc}). 
For the insulating phase, $(g\rightarrow 0)$, it follows from Eq.(\ref {gloc}) and Eq.(\ref{betagL})
that :
\begin{equation}
\lim_{g \to 0} \beta_{d}(g)\rightarrow \ln \frac{g}{g_{c}}
\label{betaloc}
\end{equation}
which is negative for $g \rightarrow 0$.
For the conducting phase (large $g$),
it follows from Eq.(\ref{gconduc}) and Eq.(\ref{betagL}) that :
\begin{equation}
\lim_{g \to \infty} \beta_{d}(g) \rightarrow  d-2
\label{betamet}
\end{equation}
which is positive for $d>2$. 
Assuming the existence of two perturbation expansions over the ``coupling'' $g$
in the limits of weak and strong ``couplings'',  one can write corrections to
Eq.(\ref{betaloc}) and Eq.(\ref{betamet}) in the following form :
\begin{eqnarray}
\label{betasmall} 
\beta_{d}(g\rightarrow 0)=\ln \frac{g}{g_{c}}(1+b_{d}g+ \cdots ) \\
\beta_{d}(g\rightarrow \infty)=d-2 - \frac{\alpha_{d}}{g} + \cdots  \qquad \alpha_{d}>0
 \label{betalarge}
\end{eqnarray}
Assuming these and a smooth monotonous $\beta_{d}(g)$, it is easy to plot the $\beta$-function
qualitatively for all $g$ and $d$, as shown in Figure \ref{fig2}. 

For $d>2$,  $\beta_{d}(g)$ must have a zero: $\beta_{d}(g_{c})=0$, where $g_{c}$ is the critical conductance.
The slope at the zero is positive, so this zero of $\beta_{d}(g)$ corresponds to an unstable fixed point of Eq.(\ref{betagL}). 
This means that a small positive or negative departure from the zero will lead  asymptotically 
to very different behaviors of the conductance. As $g$ moves from $g>g_{c}$ to $g<g_{c}$,
mobility is lost and the wavefunction is truly localized. This sharp transition is the mobility edge.

The state of a medium is supposedly determined by disorder at microscopic distances of the
order of typical spacing $a$ between neighboring sites.
Using $g (a)$ as an initial value and integrating Eq.(\ref{betagL}), it is easy to find its properties for the 2 cases : \\
$\bullet$  For $g(a) > g_{c}$, the conductivity $\sigma =g(L)L^{2-d}$ tends
for $L\rightarrow\infty$ to a constant non-zero value. \\
$\bullet$ For $g(a)<g_{c}$ in the limit of $L\rightarrow\infty$ we get
insulating behavior; that is, $\sigma =0$.  \\
We see that the behavior of $\beta_{d}(g)$ close to its zero determines
the critical behavior at the transition. There is a sharp transition for $g(a)$ larger or smaller than
$g_{c}$. 

For $g\sim g_{c}$, $\beta_{d}(g)$ is linear in $\ln g$ (see Figure \ref{fig2}), so we have the following approximation:
\begin{equation}
\beta_{d}(g)\approx \frac{1}{\nu} \ln \frac{g}{g_{c}} \approx
\frac{1}{\nu}\frac{g-g_{c}}{g_{c}}   \label{betappr}
\end{equation}
where $1/\nu$ is positive and is equal to the bracket part of Eq.(\ref{betasmall}) evaluated at $g_{c}$.
First we consider the case where $g(a)>g_{c}$. Since $\beta_{d}(g)>0$, the flow is to large $g$, so we
integrate Eq.(\ref{betagL}) to find $g(L)$.
We can approximate the integral by first integrating Eq.(\ref{betagL}) using Eq.(\ref{betappr}) until $\beta_{d}(g)$ reaches $d-2$ at $g^{*}$, then we use $\beta_{d}(g)=d-2$ (Eq.(\ref{betamet})) to reach large $L$ :
\ba
\int^{g^{*}}_{g(a)} \nu \frac{d \ln (g/g_{c})}{\ln (g/g_{c})} + \int^{g(L)}_{g^{*}} \frac{d \ln (g/g_{c})}{d-2} = \ln (L/a)
\ea
which yields an approximate
\ba
g(L) = g_{c}\left(\frac{L}{a}\right)^{d-2} \left(\ln (g(a)/g_{c})\right)^{(d-2)\nu}
\ea
We see that
the behavior of the conductivity for $L\rightarrow\infty$ :
\begin{equation}
\sigma\approx
{A}\frac{g_{c}}{a^{d-2}}\left(\frac{g(a)-g_{c}}{g_{c}}\right)^{(d-2)\nu}
\label{sigcond}
\end{equation}
where the constant $A=1$ in the above approximation, but is a function of the actual shape of $\beta_{d}(g)$.
Here $\sigma$ is independent of $L$.
The conductivity continuously goes to zero
for $g(a)\rightarrow g_{c}$,\ and $\sigma_{c}\approx 1/a^{d-2}$ is the characteristic scale of conductivity at the conductor-insulator transition.
Eq.(\ref{sigcond}) can be written as :
\begin{equation}
\sigma\approx \frac{Ag_{c}}{\xi_{c}^{d-2}}
\end{equation}
where we have introduced the conductivity length
\ba
\label{corrleng}
\xi_{c} = a\left|\frac{g(a)-g_{c}}{g_{c}}\right|^{-\nu}   
\ea
For $g>g_{c}$, $\xi_{c}$ is the correlation length that determines the behavior of the conductivity 
close to the mobility edge, when $\xi_{c} \gg l$. As $g(a) \rightarrow g_{c}$, $\xi_{c}$ blows up with critical exponent $\nu$, and $\sigma \rightarrow 0$.

Next we consider the case when the initial $g(a)<g_{c}$. Here $\beta_{d}(g)<0$, so the flow is towards $g=0$. Integrating Eq.(\ref{betagL}) with $\beta_{d}(g)$ given by
Eq.(\ref{betappr}) for $g \lesssim g_{c}$ and with $\beta_{d}(g) = \ln g/g_{c}$ for small $g$ (see Eq.(\ref{betasmall})), we obtain 
\begin{equation}
g(L)\approx g_{c}
\exp \left\{-B\left| \ln\left(\frac{g(a)}{g_{c}}\right)\right|^{\nu}\frac{L}{a}\right\}
\label{glocal}
\end{equation}
where $B$ is a constant.
Comparing this to Eq.(\ref{gloc}), we notice that
\begin{equation}
\xi \sim a\left|\frac{g(a)-g_{c}}{g_{c}}\right|^{-\nu}     
 \label{xicor}
\end{equation}
where $\nu$ is the critical exponent of localization length $\xi$. 
Now $\xi$ plays the role of the correlation length of the transition: 
it diverges at the transition. As in any critical phenomenon with a single 
correlation length, the conductivity length on the conducing side and the localization 
length on the insulating side diverge with the same exponent $\nu$. 

Next, using Eq.(\ref{betalarge}), we can solve $\beta_{d}(g_{c})=0$
to obtain :
\begin{equation}
g_{c}\simeq \frac{\alpha_{d}}{d-2}    \label{gcrit}
\end{equation}
Now we have:
\begin{equation}
\beta_{d}(g\sim g_{c})\approx (d-2)\left(\frac{g(a)-g_{c}}{g_{c}}\right)
\label{betad-2}
\end{equation}
and for the critical exponent of localization length in Eq.(\ref{xicor}), we get :
\begin{equation}
\nu \approx \frac{1}{d-2}   
\label{nuvalue}  
\end{equation}
which may be considered as the first term of the $\epsilon$-expansion from
$d=2$ (where $\epsilon=d-2$), {\em i.e.}, near ``lower critical dimension'' for
localization.

We see from Eq.(\ref{gcrit}) that $g_{c}$ depends on $\alpha_{d}$, which is clearly 
model-dependent. 
For a Fermi gas with spin-1/2 particles in 3-dimensions ($d=3$), $\alpha_{3} = \pi^{-2}$, so $g_{c} \approx 1/\pi^{2}$  \cite{Lee}. 
Here we are interested in $g_{c}$'s dependence on $d$ for large $d$. 

For tunneling in the landscape to be fast when above the transition, it would mean that the critical  
conductance $g_{c}$ for large $d \approx 10^{2}$ should be exponentially small. That is, the landscape should be in the conducting phase even for an exponentially small microscopic conductance $g(a)$. 
Noting that $g_{c} \sim 1$ for $d=3$., we see that $g_{c}$ as a function of $d$ should decrease exponentially. A glance of Figure 2 suggests that the zeros of $\beta_{d}$ (as a function of $d$) are approximately equally spaced in $\ln g$, implying that $g_{c}$ should decrease exponentially as a function of $d$. If we let 
\ba
\Delta \ln g_{c} = \ln g_{c}(d)-\ln g_{c}(d+1) = k >0
\ea
we can express $g_{c}(d)$ in terms of $g_{c}(3)$ (for $d=3$),
\ba
g_{c}(d) \simeq e^{-(d-3)k} g_{c}(3)
\ea
A simple glance at Figure 2 suggests that $k \sim 1$.
To be more specific, we like to know quantitatively the dependence of $g_{c}$ on $d$.
It is likely that this specific qualitative property is insensitive to the details of any 
particular model. In the next section, we shall examine a simple model to obtain the quantitative
dependence of $g_{c}$ on $d$.

\section{Exponentially Small Critical Conductances}

Consider a lattice model of $d$-dimensional disordered system studied by Shapiro \cite{Shapiro}. 
Since we are interested in the asymptotic behavior of $g_{c}$ as a function of $d$, where this qualitative feature should be universal, the details of the model does not concern us. We may view this model as a simple interpolation that satisfies all the general properties expected.

\subsection{A Simple Model}

The corresponding $\beta$-function in the Shapiro model is given by
\ba
\beta_{d}(g) = (d-1)  - (g+1) \ln(1 +1/g) 
\label{Shapiro1}
\ea

Recall that the critical conductance $g_{c}$ is given by the zero of the $\beta$-function. It is easy to see 
that the system has a non-trivial fixed point only for $d>2$, just as before.
For $g \rightarrow 0$, $\beta_{d}(g)  \rightarrow \ln (g/g_{c})$, reproducing Eq.(\ref{betaloc}).
 For $g \rightarrow \infty$, $\beta_{d}(g)  \rightarrow d-2$, reproducing Eq.(\ref{betamet}). The critical exponent can be expressed in terms of $g_{c}$,
\ba
\nu = \frac{1+ g_{c}}{1-(d-2)g_{c}}
\label{Shapironu}
\ea

In the $\epsilon$-expansion, for small $\epsilon=d-2 > 0$, one solves for the zero of the $\beta$-function (\ref{Shapiro1}) to obtain $g_{c} = 1/2\epsilon$. Then Eq.(\ref{Shapironu}) recovers the value of the critical exponent $\nu=1/\epsilon$ in Eq.(\ref{nuvalue}). We see that this is a simple model that reproduces the appropriate features expected.

Comparing to Eq.(\ref{gcrit}),
it gives $\alpha_{2} \approx 1/2$. For $d=3$, a numerical analysis of this model yields $g_{c} = 0.255$ and $\nu = 1.68$. This value for the critical exponent $\nu$ is higher than that in Eq.(\ref{nuvalue}) for $d=3$ from the $\epsilon$-expansion. However, this value is compatible with the estimate of $1.25 < \nu < 1.75$ \cite{Sarker}, suggesting that the leading $\epsilon$-expansion is rather crude for $\epsilon=1$.

For large $d$, we find the zero of the $\beta$-function is given by
\ba
\label{gckey}
g_{c} = e^{-(d-1)} 
\ea
that is, $g_{c}$ is exponentially small for large $d$, and
\ba
\nu \rightarrow 1
\ea
so we obtain the desired behavior for the critical conductance $g_{c}$. Comparing (\ref{gckey}) to the microscopic property (\ref{ga}) leads to the conductng/mobile condition (\ref{condition}) for large $d$.
Recall that the typical tunneling probability between 2 neighboring sites is $\Gamma_{0} \sim e^{-2a/\xi}$,
we see that the mobility condition is, for large $d$,
\ba
\Gamma_{0} > e^{-2(d-1)}
\ea
that is, an exponentially small tunneling probability can still lead to the conducting phase, that is, mobility for the wavefunction.
 
 \subsection{Disorder with Percolation}
 
This above model corresponds to a microscopically (i.e., at scales $\lesssim a$) random medium but homogeneous at a larger scale. The cosmic landscape probably looks random even at scales beyond $a$. To mimic the cosmic landscape, one may introduce additional randomness at scales larger than $a$ but still smaller than the macroscopic scale $L$. This scale can be macroscopic, or intermediate, which is some times referred to as the mesoscopic scale. Let us introduce disorder at this scale. This may be mimicked by the more general Shapiro model that actually considers localization in a macroscopically inhomogeneous medium, with percolation disorder. 
Percolation can be incorporated into the above model by introducing a probability $p$ ($1 \ge p \ge 0$) that a typical site is occupied by a random scatterer, or if a path remains open for efficient tunneling. Now, $\beta_{d}(g) \rightarrow \beta_{g}(g,p)$ becomes a function of $p$ as well,
\ba
\label{betagp}
\beta_{g}(g,p)= (d-1)  \left( 1 + \frac{1-p}{p} \ln(1-p) \right) - (g+1) \ln(1 +1/g)
\ea
where the running of $p$ is given by 
\ba
\label{betapp}
\beta_{p} (p) =\frac{\partial p}{\partial \ln L} = p \ln p -(d-1) (1-p) \ln (1-p) 
\ea
In this simple model, we see that $\beta_{p}$ is independent of $g$.
Here, it is easy to see that $p=1$ is a fixed point since $\beta_{p} (p=1)=0$. At this fixed point, $\beta_{g}(g,p)$ (\ref{betagp}) reduces to the above $\beta$-function (\ref{Shapiro1}). In general, $g_{c}$ is a now a function of both $p$ and $d$. For any given $p$, we see that $\beta_{g}(g,p)$ is simply given by $\beta_{d}(g)$ (\ref{Shapiro1}) with $d$ replaced by an effective $d_{p}$ which is a function of $p$,
\ba
d_{p}  = 1 + (d-1)\left( 1 + \frac{1-p}{p} \ln(1-p) \right)
\ea
so, for large $d$, we now have 
\ba
\label{conditionp}
g_{c} (p) = e^{-(d_{p}-1)} 
\ea 
That is, the condition (\ref{condition}) for fast tunneling now becomes
\ba
d_{p} > \frac{a}{\xi} +1
\ea
and the critical exponent $\nu$ for $\beta_{g}$ becomes
\ba
\nu_{g} = \frac{1+ g_{c}}{1-(d_{p}-2)g_{c}}
\label{Shapironudp}
\ea

Let us consider the zeros of $\beta_{p} (p)$ at $p=p_{c}$, where $p_{c}$ satisfies
\ba
p_{c} \ln p_{c} =(d-1)(1-p_{c}) \ln(1-p_{c})
 \ea
For $d \ge 2$, there are 3 fixed points, which we shall label as $p_{1}=0$, $p_{2}$ and $p_{3}=1$ : 
besides the 2 fixed points, namely $p_{1}=0$ and $p_{3}=1$, there is another fixed point at $p=p_{2}$ 
in between; that is, $p_{3}> p_{2} >p_{1}$ or $ 1> p_{2} >0$. 
For $d=2$, it is easy to see that $p_{2}=1/2$. For $d=3$, Ref.\cite{Shapiro} gives $p_{2}=0.16$.
For large $d$, we have
\ba
\label{p2d}
p_{2} \simeq  e^{-(d-1)}
\ea
Expanding $\beta_{p} (p)$ around $p_{c}$, we have, for $|\delta =  p - p_{c}| \ll 1$, 
\ba
\beta_{p} (p) \approx \delta \left( d + \frac{\ln p_{c}}{1-p_{c}} \right) = \frac{\delta}{\nu_{p}}
\ea
where the critical exponent $\nu_{p}$ for $p$ is determined in terms of $p_{c}$.
For large $d$, we see that $p_{2}$ is an unstable fixed point while $p_{1}=0$ and $p_{3}=1$ 
are stable fixed points. Using (\ref{p2d}), we have, for large $d$,
\ba
\nu_{p}(p_{2}) \approx 1
\ea
That is, for $p < p_{2}$, $p \rightarrow 0$, while for $p > p_{2}$, $p \rightarrow 1$.
We see that $d_{p} =1$ at the fixed point $p_{1}=0$.
For such a small $d_{p}$, the medium is insulating. So one may conclude that the medium is insulating for $p <p_{2}$.
Following Eq.(\ref{p2d}), we see that the attractive region $p_{2}> p \ge 0$ for the insulating phase is exponentially small. 

For $p>p_{2}$, depending on the initial microscopic values of $g$ and $p$, the medium is in an insulating phase  with $g \rightarrow 0$ when $g(p) < g_{c}(p)$, or, in a conducting phase when $g(p) > g_{c}(p)$ : 
$\beta (g) \rightarrow d-2$, where $g_{c}(p)$ is given by Eq.(\ref{conditionp}).
Since $p_{2} \rightarrow 0$ as $d \rightarrow \infty$, we see that 
the condition for the conducting/mobile phase is quite insensitive to disorder at the mesoscopic scale.

We see that there are universality properties that one may extract from the cosmic landscape that is most likely insensitive to any of the details. However, a better knowledge of its structures can be important for finding out which universality class the renormalization group properties of the cosmic landscape belongs to.  

\section{A Critical Cosmological Constant}

Now we like to see how a critical $\Lambda_{c}$ can emerge. It is reasonable to assume that tunneling takes place from a site with $\Lambda_{1}>0$ only to any other sites with $\Lambda$ which is smaller but semi-positive, that is, $\Lambda_{1} \ge \Lambda  \ge 0$. For the discussion to be applicable to general uncompactified spatial dimensions, let $\hat \Lambda$ designate the mass scale of the vacuum energy density, so the usual vacuum energy density goes like $\Lambda={\hat\Lambda}^{4}$ in 4-dimensional spacetime.

Let us start with the simple condition for mobility (\ref{condition}), where, for large $d$, 
\ba
\label{cond2}
d(\Lambda) >  \frac{a(\Lambda)}{\xi (\Lambda)} \quad \rightarrow  \quad d >  \frac{a(\hat \Lambda)}{\xi}
\ea
where, to simplify the problem, we assume that $d$ and $\xi$ are insensitive to $\Lambda$. 
As $\Lambda$ decreases, a site has fewer nearby sites to tunnel to, so the effective site 
spacing $a$ increases. As $a$ increases, at certain critical value, the above condition is no longer satisfied and true localization takes over. (We expect that effectively $d$ decreases, but we ignore this effect here. Including this effect changes only the quantitative picture.) 

Suppose sites with different $\Lambda$ are irregularly distributed in the landscape. Let there be $N_{T}$ classically stable sites with $M_{s} > \Lambda \ge 0$ in a region in the moduli space with size $L$.
If we start at a site with $\hat \Lambda=M_{s}$, then the typical spacing $a$ is given by
\ba
N_{T} =N(M_{s}) = \left(\frac{L}{a(M_{s})}\right)^{d}
\ea
Let $f(\hat \Lambda)$ be the fraction of sites with semi-positive vacuum energy $\hat \Lambda< M_{s}$, with $f(M_{s})=1$. (That is, the distribution function is given by $N_{T} (df/d\Lambda)$.) 
Then the number of sites in this region with vacuum energy smaller than $\Lambda$ is given by
\ba
N(\hat \Lambda) = f(\hat \Lambda) N_{T} = \left(\frac{L}{a(\hat \Lambda)}\right)^{d}
\ea
or
\ba
a(\hat \Lambda) = a(M_{s}) f(\hat \Lambda)^{-1/d}
\ea
Since $f < 1$ for $\hat \Lambda< M_{s}$, we see that $a(\hat \Lambda) > a(M_{s})$,
and $a$ increases as $\hat \Lambda$ decreases. Following Eq.(\ref{cond2}), we see that there is a critical $\Lambda_{c} = {\hat \Lambda}_{c}^{4}$, below which mobility stops and the wavefunction is truly localized. The universe lives exponentially long at any low $\Lambda$ site, where $\Lambda < \Lambda_{c}$.

As an illustration, one may take $f(\hat \Lambda) \approx (\hat \Lambda/M_{s})^{s}$, so $a(\hat \Lambda)/a(M_{s}) \approx   (\hat \Lambda/M_{s})^{-s/d}$. Taking $a(M_{s}) \sim 1/M_{s}$ and similarly $\xi \sim 1/M_{s}$, we have
\ba
{\hat \Lambda}_{c} \approx d^{-d/s} M_{s} 
\ea
Since the cosmological constant $\Lambda$ in 4-dimensional spacetime goes like ${\hat \Lambda}^{4}$, $s=4$ corresponds to a flat $\Lambda$ distribution. For this distribution, the critical 4-dim. cosmological constant goes like
\ba
  \Lambda_{c} \sim d^{-d}M_{s}^{4} 
\ea
so $d \sim 60$ looks quite reasonable.
(Recall that ${\hat \Lambda}_{c} /M_{s} \sim 10^{-30}$ to $10^{-20}$.) 
In fact, a smaller $s$ or a larger $d$ will lead to a much too small $\Lambda_{c}$. 
We see that the vastness of the cosmic landscape, as parameterized by the large $d$, is crucial to this particular approach to the cosmological constant problem.

 \section{Discussions}
 
As we have discussed, we like to avoid eternal inflation totally in the history of our universe.
If the universe goes through inflation for more than one e-fold in a $\Lambda$ meta-stable site, then eternal inflation is unavoidable, since tunneling of some patches of the inflating universe will still leave other patches to continue inflating.
At a given site, the time to go through one e-fold of inflation is a Hubble time, i.e., $\Delta t =H^{-1}\sim M_{Planck}/ {\hat \Lambda}^{2}$ (recall that we have defined $\hat \Lambda$ to be the scale of the cosmological constant). (If the universe at that site starts with some radiation, then it will take additional time to reach the inflationary stage. So the time the universe can stay in a $\Lambda$ meta-stable site without eternal inflation may be longer than the Hubble time. Let us ignore the presence of radiation for the moment.)
In the absence of resonance tunneling, the time $\delta t$ that the universe stays at the $\Lambda$ site is $\delta t \sim 1/H \Gamma_{0} \gg 1/H$, so eternal inflation is unavoidable.

To avoid eternal inflation in the quantum landscape, fast tunneling has to be faster for larger $\Lambda$. 
In the mobility phase, we can crudely estimate the time $\delta t$ it takes for the universe to move from one site to another, using the simple formula :  potential difference $V$ equals the product of the current and the resistance $\sim1/g$. To get an order of magnitude estimate, we let $V \sim \hat \Lambda$, the current $\sim (a/\delta t)/L$ and $g(L) =\sigma L^{d-2} \sim B (L/a)^{d-2}$ where $\sigma$ is the rescaled conductivity and $B$ is a dimensionless finite constant.
This yields the time it takes the state to move from a site to a neighboring site
$\delta t \sim (a/L)^{d-1}(1/{\hat \Lambda})$, or
\ba
\delta t < 1/{\hat \Lambda} < \Delta t  \nonumber
\ea
So we see that the avoidance of eternal inflation in the early universe is probably automatic when the universe is in the conducting component of the landscape, i.e., when $\Lambda >\Lambda_{c}$.
So it is reasonable to assume that no eternal inflation happens during the mobile phase.

Once the universe reaches an insulating vacuum site,
its lifetime there is longer than the critical lifetime $T_{c}$,
\ba
T_{c} \sim \frac{1}{M_{s}\Gamma_{0}} =  \frac{1}{M_{s}}  e^{2(d-1)} 
\ea
With $d \sim 60$ and $M_{s}^{-1} \sim 10^{-30}$ sec., we see that $T_{c}$ is rather close to the age of our universe (within a few orders of magnitude). So the sharp transition from mobility (fast enough not to allow eternal inflation) to being trapped at a site with an exponentially long decay time seems easily attainable. 

One can envision the following scenario, where the universe is spontaneously created from nothing
\cite{Vilenkin:1982de,Hartle:1983ai} (i.e., no classical spacetime) at a specific site (or a neighborhood region) in the cosmic landscape with $\Lambda$ somewhat below the string scale \cite{Firouzjahi:2004mx,Sarangi:2005cs,Brustein:2005yn,Sarangi:2006eb,Barvinsky:2006uh}. Fast tunneling enables mobility in the cosmic landscape, and the wavefunction of the universe moves rapidly towards sites with smaller $\Lambda$. This motion at times can be classical, leading to inflation (e.g., in the $I$ region in Figure 3). Afterwards, it will continue to move towards lower $\Lambda$ sites until it reaches a site with $\Lambda <\Lambda_{c}$ where it has been trapped till now.

\begin{figure}[ht]
\begin{center}
\leavevmode
 \includegraphics[width=0.7\textwidth,angle=0]{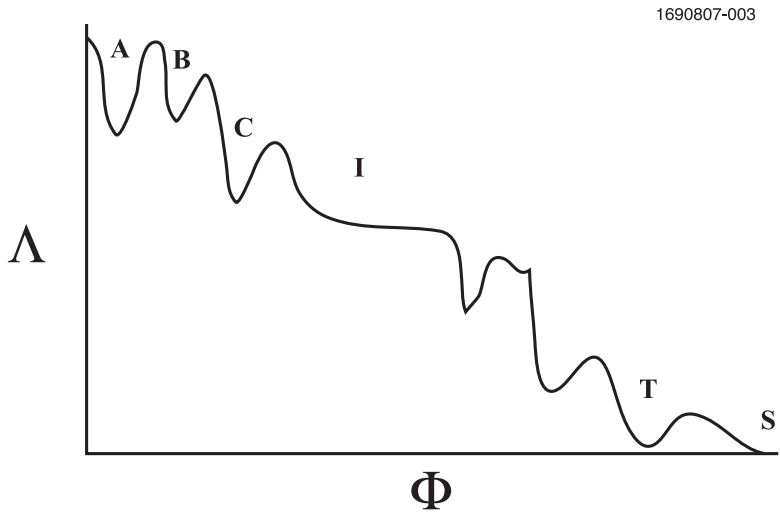} 
\caption{A cartoon sketch of the landscape. Suppose the universe is created at site $A$. Fast tunneling brings it to $I$, where inflation takes place as it rolls down the slope there. After inflation, it continues towards $T$ when it stays there for an exponentially long time before moving to a supersymmetric (and probably decompactified) site $S$. $T$ can be our today's universe.
}
\label{fig3}
\end{center}
\end{figure}

It is interesting to speculate more about what happens after inflation.
Towards the end of inflation, the inflaton field (or the associated
tachyon mode) rolls down to the bottom of the inflaton potential. The
universe is expected to heat up to start the hot big bang epoch.
It is not clear which phase the wavefunction belongs to (conducting or insulating) if the energy density (radiation plus vacuum energy) is bigger than the critical value but the vacuum energy itself is below the critical value. This is a physics question that should have a definitive answer. Before this issue is resolved, let us entertain both possibilities.

If it is in the insulating phase, then that long-living meta-stable vacuum state should have an exponentially small vacuum energy (otherwise, it will continue to move down). It can be our universe. 

Now let us assume that we are still in the conducting phase when 
the total energy (radiation plus vacuum energy) density is bigger than the critical value even though the vacuum energy density itself is below the critical value \cite{Sarangi:2005cs}. In this case,
the wavefunction of the universe can spread over a large range,
a linear superposition of many (positive cosmological constant) vacua (say, $N$) at the
foothill of the inflaton potential. This is like a Bloch wavefunction
as proposed in Ref.\cite{Kane:2004ct}. 
The ground state energy of this wavefunction is
expected to be much smaller than the average vacuum energy (by a factor of  $N^{-1}$),
a property of Bloch wavefunctions. As the universe evolves, tunneling among the
vacua will become suppressed, due to cooling as well as decoherence.
As energy density decreases, sites with larger vacuum energy densities begin to decouple.
Eventually, the wavefunction collapses to a single vacuum.   Since the Bloch wavefunction is able to sample many vacua, and it already has a very small energy density, it is natural to expect that it will collapse to the vacuum state with the
smallest positive vacuum energy within the sample. This may partly
explain why the dark energy is so small. The question is whether the collapse of 
the wavefunction is fast and efficient enough to preserve big bang nucleosynthesis.
The collapse of the wavefunction is a consequence of decoherence, where it is natural to treat the radiation/matter as the environment and the $\Lambda$ vacuum as the system. So, even if gravity does not distinguish radiaton/matter from the vacuum energy well enough, quantum mechanics may.

One can speculate further.
Although the scaling theory naively implies that there is no transition in $d=2$ (see Figure 2), it is only marginally so. A small correction can change this qualitative picture, as it happens in some systems in condensed matter physics. So far, we have considered the transition between insulation and conductivity. In condensed matter physics, there is also a transition between insulation and superconductivity. Can this happen in the landscape ? This is an interesting question. Suppose the naive dimensional scaling $g \approx L^{d-2}$ is modified by an anomalous dimension so
 $g \approx L^{d-2 + \delta}$, with a positive $\delta$, then the conductivity will grow with $L$. With the conductivity approaching infinity, the landscape can actually be superconducting. In this case, it is possible that tiny differences in properties between vacua (say, 2 vacua with the same total energy density but one with a slightly smaller $\Lambda$ and more radiation) is enough for the wavefunction to flow there and then localize there. The interplay between coherence and decoherence should play a crucial role here. Hopefully, this paper provides new directions to tackle this fundamental problem.

 It is also possible that inflation happens during fast tunneling \cite{Davoudiasl:2006ax,Freese:2006fk,Allahverdi:2007wh,Huang:2007ek}. Eventually, the wavefunction arrives at a site with a small $\Lambda$, where $\Lambda$ is smaller than the critical $\Lambda_{c}$. Since tunneling from that site is exponentially long, the wavefunction is truly localized there. This new type of inflation should be studied carefully. In summary, inflation can happen before, during and/or after fast tunneling. There are numerous inflationary scenarios one can consider. Hopefully the solution to the cosmological constant problem imposes tight constraints on the inflationary scenario. 

 \section{Remarks}

That today's dark energy is positive but exponentially smaller than the typical scale of gravity is at first sight very mysterious. Our universe is sitting at a stable vacuum state or a meta-stable vacuum state that has lived for more than 13 billion years. A necessary condition for string theory to describe nature is to have such a vacuum state solution. The vastness of the cosmic landscape, with exponentially many vacua solutions, allows for such a small $\Lambda$ site to exist. In Ref.\cite{Tye:2006tg} and this paper,
we argue that the same vastness of the landscape offers a dynamical reason why our universe should end up in a very small positive $\Lambda$ site, without having to go through eternal inflation. In this sense, string theory may provide a resolution to one of the deepest questions in physics. 

A detailed understanding of the stringy landscape will allow us to get a more precise set of renormalization group equations. As discussed earlier, we expect $\Lambda_{c}$ to be exponentially small compared to the string scale, and today's dark energy must be smaller than $\Lambda_{c}$. So it is important to obtain a reliable estimate of $\Lambda_{c}$.  A better knowledge of the landscape will certainly help. 

In condensed matter physics, the transition from high-temperature superconductor to an insulator can be sharp. Furthermore, in some systems, superconductivity can take place even in the region of localization. This is a very active area of research. We hope to learn more from the interplay between string theory and condensed matter physics.

\vspace{8mm}

I thank Tomas Arias, Tom Banks, Raphael Bousso, Piet Brouwer, Rong-gen Cai, Xingang Chen, Ed Copeland, Chris Henley, Andre LeClair, Miao Li, Liam McAllister, Sash Sarangi, Gary Shiu, Ben Shlaer and Jiajun Xu for discussions. The revision was prepared while visiting Kavli Institute for Theoretical Physics China, Beijing. I thank Yue-liang Wu for his hospitality.  This work is supported by the National Science Foundation under grant PHY-0355005.

\vspace{8mm}

\appendix

\section{Resonance Tunneling}

Here resonance tunneling pertinent to Ref.\cite{Tye:2006tg} and this paper is briefly reviewed. The definitions of efficient tunneling and fast tunneling used in Ref.\cite{Tye:2006tg} are clarified.

In quantum mechanics or in quantum field theory, the tunneling probability (transmission coefficient) $T_{A \rightarrow B}$ from the meta-stable site $A$ to $B$ over a barrier between them is in general exponentially suppressed (see Figure 3). Naively, tunneling from $A$ to $C$ is expected to be doubly suppressed, that is, 
\ba
\label{naive}
T_{A \rightarrow C} \simeq T_{A \rightarrow B} T_{B \rightarrow C} 
\ea
unless the energy of the wavefunction coincides with an energy eigenvalue for a state localized in $B$. This is the resonance condition. When that happens, tunneling in quantum mechanics is exponentially enhanced. This is known as resonance tunneling, where 
\ba
\label{restun}
T_{A \rightarrow C}  = \frac{4T_{A \rightarrow B} T_{B \rightarrow C}}{(T_{A \rightarrow B} + T_{B \rightarrow C})^{2}} 
\ea
Note that resonance tunneling requires a quantum treatment at the energetically allowed region $B$.
Since $T_{A \rightarrow B}$ and $T_{B \rightarrow C}$ in general can differ by orders of magnitude,
$T_{A \rightarrow C}$ may still be exponentially small.

However, in the landscape, we do not expect the wavefunction to have precisely the resonance eigenvalue. In general it will have a spread and the probability of hitting the resonance eigenvalue is
given by the resonance width divided by the spacing between resonances,
\ba
P(A \rightarrow C) = \frac{1}{2 \pi} (T_{A \rightarrow B} + T_{B \rightarrow C})
\ea
So the average tunneling probablity is
\ba
\label{averest}
<T_{A \rightarrow C}> = P(A \rightarrow C) T_{A \rightarrow C} \sim  
\frac{T_{A \rightarrow B} T_{B \rightarrow C}}{T_{A \rightarrow B} + T_{B \rightarrow C}} 
\ea
which, to a good approximation, is given by the smaller of the 2 tunneling probabilities. This is exponentially bigger than the naive expectation (\ref{naive}).

Efficient tunneling happens when $T_{A \rightarrow B} \rightarrow T_{B \rightarrow C}$. Now, efficient tunneling can become fast tunneling (i.e., tunneling probability approaching unity) when 
 the wavefunction has precisely the resonance eigenvalue;
 as $T_{A \rightarrow B} \rightarrow T_{B \rightarrow C}$, resonance tunneling (\ref{restun}) leads to
\ba
T_{A \rightarrow C}  \rightarrow 1
\ea
We refer to this as fast efficient tunneling. 
The decay of $A$ goes to $C$ exclusively, even when there are other sites around. 
However, we do not expect this to be the case for the wavefunction of the universe in the cosmic landscape. (Most likely the wavefunction has a spread in eigenvalues.)
Instead, we expect the average tunneling probability (\ref{averest}) to apply, that is, $<T_{A \rightarrow C}>  \sim T_{A \rightarrow B} = T_{B \rightarrow C}$
for efficient tunneling. The above discussion can be extended to multi-step tunneling.

Now, fast tunneling can also happen when the wavefunction is in a $d$-dimensional landscape, with large $d$, as suggested by Eq.(\ref{rst}). In this case, the tunneling probabilities from $A$ to more than one site in its neighborhood can be comparable. They can happen via normal tunneling, resonance tunneling and/or efficient tunneling. In the case of a random potential or the cosmic landscape, presumably a specific tunneling channel (or a small subset) dominates. This approximation leads naturally to a disorder and/or percolation picture.

\section{Resonance Tunneling Effect in Quantum Field Theory}

The above resonance tunneling effect is discussed in the context of quantum mechanics.
It is important to know whether and how it happens in quantum field theory and in gravity. Although some argument for its presence is given in Ref.\cite{Tye:2006tg}, this remains an open question at this moment \cite{Sarangi:2007jb,Copeland:2007qf}. 
For the cosmic landscape, we actually need only the averaged form
(\ref{averest}). Here we like to justify this formula in 2 ways. In this appendix, we give a simple reasoning
of its origin independent of resonance tunneling. In the next appendix, we argue how it arises in Hawking-Moss tunneling, which is tunneling in quantum field theory in the presence of gravity.

  Intuitively, the meaning of the formula (\ref{averest}) is actually very simple \cite{Chen:2006ni}. 
The tunneling rate is given by
$\Gamma_{i \rightarrow j} \simeq H T_{i \rightarrow j}$. 
So the time it takes to tunnel from ${A \rightarrow C}$, namely $t (A \rightarrow C)$, is equal to the time it takes to tunnel from $A \rightarrow B$ ($=1/\Gamma_{A \rightarrow B}$) plus the time it takes to tunnel from $B \rightarrow C$ ($=1/\Gamma_{B \rightarrow C}$),
\ba
t (A \rightarrow C) = t (A \rightarrow B) + t (B \rightarrow C)
\ea
 Since this total tunneling time gives the inverse of the tunneling rate $\Gamma_{A \rightarrow C}$, we have
\ba
\label{time}
\frac{1}{T_{A \rightarrow C}}  = \frac{1}{T_{A \rightarrow B}}  + \frac{1}{T_{B \rightarrow C}} 
 \ea
which, for a generic wavefunction, is simply Eq.(\ref{averest}), with the averaging implied.
Notice that, to a good approximation, $T_{A \rightarrow C}$ (\ref{time}) is given by
\ba
\label{obvious}
T_{A \rightarrow C} \approx {\rm Minimum} \left( T_{A \rightarrow B}, T_{B \rightarrow C} \right)
\ea
For $T_{A \rightarrow B} = T_{B \rightarrow C} = \Gamma_{0}$, we see that $T_{A \rightarrow C} \sim \Gamma_{0}$, not $\Gamma^{2}_{0}$.

\section{Hawking-Moss Tunneling}

As mentioned just now, for the landscape, we actually use only the simple form (\ref{time}) or
(\ref{obvious}) in the presence of gravity. 
Here, I give a brief argument for its justification within the Hawking-Moss (HM) tunneling mechanism \cite{Hawking:1981fz}. For $a> \xi$, the barrier is probably rather broad and the thin-wall approximation (the Coleman-De Lucia bounce) probably breaks down, but the HM tunneling formula should be applicable.

Consider the tunneling (from $A$ to $B$ in Figure 3) probability per unit volume in a single scalar field theory in 4 dimensions. Let $\phi_{A}$ be the position of the classical minimum of the potential $V(\phi)$ at site $A$,
with value $V(\phi_{A})$, and $\phi_{AB}$ be the position of the maximum of the potential $V(\phi)$ between $A$ and $B$, with value $V(\phi_{AB})$. In the HM tunneling, the tunneling probability per unit volume is given by ($G_{N}=M_{Pl}^{-2}$)
\ba
\label{HMT1}
T_{A \rightarrow B} \sim e^{-S_{E}(\phi_{AB}) +  S_{E}(\phi_{A})} = \exp \left(- \frac{3 M_{Pl}^{4}}{8V(\phi_{A})} + \frac{3 M_{Pl}^{4}}{8V(\phi_{AB})} \right)
\ea
A proper interpretation of the HM tunneling is, I believe, given in Ref.\cite{Starobinsky:1986fx,Goncharov:1987ir,Linde:1990}. During each Hubble time period $1/H$, $\phi$ experiences quantum fluctuation with amplitude $\delta \phi \sim H/2 \pi$ and horizon length scale $\sim 1/H$. These quantum jumps lead to local changes in $\phi$, so that $\phi$ can stochastically climb up the potential 
from the minimum $V(\phi_{A})$ to the top $V(\phi_{AB})$. Once $\phi$ is at the top of the potential, it can simply roll down (classically) to $B$. Since the rolling down happens with probability of order unity, $T_{A \rightarrow B}$ measures the probability to find $\phi$ at the top and 
at $B$ and so has a simple interpretation as the fraction of the comoving volume of the universe in $B$. The spatially homogeneous HM bounce is particularly good if the potential barrier is broad. 

Now, consider the tunneling from $A \rightarrow C$ (see Figure 3). We see that once $\phi$ is at the top of the potential between $A$ and $B$, and if the top $V(\phi_{BC})$ between $B$ and $C$ is lower than
$V(\phi_{AB})$, then $\phi$ can simply roll all the way to $C$, that is
$T_{A \rightarrow C} \sim T_{A \rightarrow B}$. (Here we ignore the expansion of the universe, which provides a damping effect that should be taken into account.)

If, on the other hand, suppose $V(\phi_{BC}) > V(\phi_{AB})$. 
After reaching $\phi_{AB}$, $\phi$ then classically rolls 
from $\phi_{AB}$, passes $B$ and rolls up the potential between $\phi_{B}$ and $\phi_{BC}$ to reach the height of $V(\phi_{AB})$. Now, quantum fluctuation has to take over to move $\phi$ up to the top of the potential $V(\phi_{BC})$ at $\phi_{BC}$. The probability per unit volume for achieving this is given by  
\ba
\label{BCA}
T(\phi_{BC}) =\exp \left(- \frac{3 M_{Pl}^{4}}{8V(\phi_{AB})} + \frac{3 M_{Pl}^{4}}{8V(\phi_{BC})} \right)
\ea
so together, using Eqs.(\ref{HMT1}, \ref{BCA}),
\ba
\label{HMT2}
T_{A \rightarrow C} \sim T_{A \rightarrow B} T(\phi_{BC}) = \exp \left(- \frac{3 M_{Pl}^{4}}{8V(\phi_{A})} + \frac{3 M_{Pl}^{4}}{8V(\phi_{BC})} \right)
\ea
So, in general, we see that $T_{A \rightarrow C}$ is given by the smaller (minimum) of (\ref{HMT1}) and (\ref{HMT2}), that is, it is dictated by the larger of $V(\phi_{BC})$  or $V(\phi_{AB})$.
If $V(\phi_{A}) = V(\phi_{B})$, we see that 
\ba
\label{resHM}
T_{A \rightarrow C} \simeq {\rm Minimum} \left( T_{A \rightarrow B}, T_{B \rightarrow C} \right)
\ea
in agreement with $T_{A \rightarrow C}$ (\ref{obvious}) obtained via resonance tunneling effect or intuitive considerations.

It is easy to extend this formula to the multi-step case, that is, the tunneling probablity depends only on the initial false vacuum value $\phi_{A}$ and the value of the highest point of the barrier between and $A$ and the true vacuum  (or the final vacuum of interest), so that transition to distant parts of the landscape can be as likely as to nearby regions. (Damping effects such as the expansion of the universe will provide a cut-off.)
Note that the picture here is close to Ref.\cite{Clifton:2007en} but not to Ref.\cite{Weinberg:2006pc}.
Of course, strcitly speaking, HM tunneling is not a tunneling process (more like a random walk) and Eq.(\ref{obvious}) or (\ref{resHM}) is not due to resonance tunneling. Nevertheless, it is reassuring that the important end result relevant in this paper and in Ref.\cite{Tye:2006tg} is the same.

\end{document}